\documentclass[10pt,aps,prl,twocolumn,showpacs,superscriptaddress,nofootinbib,nobibnotes,longbibliography,floatfix]{revtex4-1}

\usepackage[utf8]{inputenc}
\usepackage[T1]{fontenc}
\usepackage{comment}
\usepackage{bm}
\usepackage[normalem]{ulem}\usepackage{mathtools,amsmath,amssymb,amsfonts,mathrsfs,eucal,graphicx,tensor,csquotes,accents,commath,chngcntr,siunitx}
\usepackage[dvipsnames]{xcolor}
\usepackage[unicode]{hyperref}
\hypersetup{colorlinks=true, citecolor=MidnightBlue,
            linkcolor=MidnightBlue, urlcolor=MidnightBlue, linktocpage=true}
\usepackage[normalem]{ulem}

\DeclareMathAlphabet{\mathpzc}{OT1}{pzc}{m}{it}
\definecolor{darkgreen}{rgb}{0.0, 0.6, 0.0}

\begin{document}

\date{\today}

\title{Probing Spacetime Symmetries Using Gravitational Wave Ringdown}

\author{Rajes Ghosh}
\email{rajes.ghosh@icts.res.in }
\affiliation{International Centre for Theoretical Sciences, Tata Institute of Fundamental Research, Bangalore 560089, India.}
\author{Akash K Mishra}
\email{akash.mishra@saha.ac.in}
\affiliation{International Centre for Theoretical Sciences, Tata Institute of Fundamental Research, Bangalore 560089, India.}
\affiliation{Theory Division, Saha Institute of Nuclear Physics, 1/AF Bidhan Nagar, Kolkata 700064, India}
\author{Sudipta Sarkar}
\email{sudiptas@iitgn.ac.in}
\affiliation{Indian Institute of Technology, Gandhinagar, Gujarat 382355, India.}

\begin{abstract}

The uniqueness and rigidity theorems assert that the asymptotically flat, vacuum, stationary rotating black hole solution in general relativity must be the Kerr solution, exhibiting novel symmetries such as axisymmetry and circularity. In our analysis of post-merger ringdown signal from coalescing black hole binary systems, we identify potential observational signatures for deviations from these Kerr symmetries. Utilizing ringdown data from the gravitational wave event \texttt{GW150914}, we place significant constraints on such deviations. Our analysis introduces a new and novel approach for testing spacetime symmetries through gravitational wave observations.

\end{abstract}
\maketitle

\noindent{\bf{\em Introduction.}}
Gravitational waves (GWs) have been at the forefront of modern cutting-edge research since their first observation \cite{LIGO1, LIGO2, LIGOScientific:2020ibl, KAGRA:2021vkt} by the LIGO and Virgo detectors~\cite{LIGOScientific:2014pky, VIRGO:2014yos}. Thanks to their ability to reveal fundamental aspects of gravity across different scales, they offer us several powerful tools to probe the yet-alluring strong field signatures of General Relativity (GR) \cite{LIGOScientific:2016lio, LIGOScientific:2020tif, LIGOScientific:2021sio, Berti:2015itd, Berti:2018vdi}. One such promising tool is black hole (BH) spectroscopy used to extract remnant quasinormal modes (QNMs) during the ringdown phase of binary mergers \cite{Regge:1957td, Vishveshwara:1970, Zerilli:1970se, Teukolsky:1973ha, Dreyer:2003bv, Detweiler, Berti:2005ys, Gossan:2011ha}. This novel method is extraordinarily successful in investigating the nature of post-merger compact objects \cite{Cardoso:2019rvt, Maggio:2020jml, Maggio:2021ans, Maggio:2023fwy} and extracting key remnant properties such as mass and spin~\cite{Carullo:2019flw, Isi:2021iql, Cotesta:2022pci, Siegel:2023lxl, Correia:2023ipz}. Furthermore, it plays a crucial role in investigating potential deviations from GR predictions~\cite{Meidam:2014jpa, Carullo:2018sfu, Isi:2019aib, CalderonBustillo:2020rmh, Carullo:2021dui, Westerweck:2021nue, Carullo:2021oxn, Dey:2022pmv, Silva:2022srr, Franchini:2023eda, Pacilio:2023mvk, Destounis:2023ruj, DAddario:2023erc, Baibhav:2023clw, Mishra:2023kng, Cano:2024jkd, Cano:2024ezp, Crescimbeni:2024sa}, which could arise from higher-curvature modifications, nontrivial environmental effects, or any other new physics.

There are a few independent ways in which these beyond-GR effects could manifest themselves. Among these possibilities, one particularly well-trodden direction is the so called \textit{post-Kerr paradigm} that introduces new parameters (hairs) in the Kerr metric and constrains them from observations~\cite{Gair:2007kr, Johannsen:2011dh, Johannsen:2013szh, Rezzolla:2014mua, Konoplya:2016jvv}. These extra hairs are typically incorporated through phenomenological spacetimes, such as the Johannsen-Psaltis~\cite{Johannsen:2011dh, Johannsen:2013szh} and Konoplya-Rezzolla-Zhidenko metrics~\cite{Rezzolla:2014mua, Konoplya:2016jvv}, which are among the widely used ones. However, this method has a significant limitation: Most post-Kerr metrics are constructed retaining all the symmetries of the Kerr solution, including stationarity ($t \to t+ dt$), axisymmetry ($\varphi \to \varphi + d\varphi$) and circularity (simultaneous reflections of $(t, \varphi) \to (-t, -\varphi)$). While this assumption leads to obvious mathematical simplifications, it fails to capture the most general scenarios that could arise from deviations from GR. In principle, beyond-GR effects might cause departures from various Kerr-symmetries and detecting such deviations can have important observational consequences \cite{Destounis:2021mqv, Destounis:2021rko, Fransen:2022jtw, Chen:2022lct, Chen:2023gwm, Ghosh:2024arw}. Moreover, even in the theoretical side, it may lead to far-reaching implications in numerous fundamental results in GR, like uniqueness and no-hair theorems that presupposes those symmetries \cite{Israel:1967wq, Geroch:1970cd, Carter:1971zc, Bekenstein:1971hc, Bekenstein:1972ky, Hansen:1974zz, Robinson:1974nf, Robinson:1975bv}. Thus, it is absolutely crucial to verify whether astrophysical spacetimes actually possess such isometries.

Hence, we can ask a well-motivated and important question: Is there a general methodology to probe departures from various Kerr symmetries? To derive an answer, we may take hints from our knowledge of standard classical/quantum mechanical systems. In those systems, it is a well known fact that \textit{symmetries always lead to degeneracies in certain physical observables}. The same principle also holds for gravitational systems like BHs. In particular, as a direct consequence of stationarity, axisymmetry and circularity, the time-domain ringdown signals of Kerr BHs can be modelled as a linear superposition of ``twin'' modes, which also showcase certain degenaracies~\cite{Berti:2005ys, Krivan:1997hc, Dorband:2006gg, Giesler:2024hcr}, 
\begin{equation} \label{ringdown}
    \begin{split}
        h(t) = \sum_{\ell, m, n} & \left\{ \mathcal{A}_{\ell m n}\, e^{\mathrm{i}\left(\omega_{\ell m n} t+\phi_{\ell m n}\right)}\, e^{-t / \tau_{\ell m n}}\, S_{\ell m n}(\theta, \varphi)\, +\right. \\
       &\left. \mathcal{A}'_{\ell m n}\, e^{\mathrm{i}\left(\omega'_{\ell m n} t+\phi'_{\ell m n}\right)}\, e^{-t / \tau'_{\ell m n}}\, S^*_{\ell m n}(\theta, \varphi) \right\}.
    \end{split}
\end{equation}
The real part of QNM frequencies and the corresponding damping times of these twin modes are related by\footnote{There are similar degenaracies in the amplitude $\mathcal{A}_{\ell m n}$ and phase $\phi_{\ell m n}$ too for the twin modes. In fact, if the initial perturbation has the equatorial reflection symmetry, then $\mathcal{A}'_{\ell m n}\, e^{\mathrm{i} \phi'_{\ell m n}} = (-1)^\ell\, \mathcal{A}_{\ell m n}\, e^{-\mathrm{i} \phi_{\ell m n}}$~\cite{Faye:2012we, Blanchet:2013haa, Capano:2021etf}, which we shall assume. Moreover, we have redefined the amplitudes w.r.t Ref.~\cite{Berti:2005ys} by absorbing an overall $M/r$ factor.}: $\omega'_{\ell m n} = -\omega_{\ell m n}$, and $\tau'_{\ell m n} = \tau_{\ell m n}$ for any fixed $(\ell, m, n)$, respectively~\cite{Berti:2005ys, Krivan:1997hc, Dorband:2006gg, Giesler:2024hcr}. One must not interpret these relations as degenaracies between the prograde and retrograde modes---their frequencies and damping time are completely different. For aligned-spin binaries, the smaller-frequency retrograde modes are typically not excited within the LIGO band~\cite{Berti:2005ys, Lim:2019xrb, Li:2021wgz}. Then, only for the twin modes at fixed $(\ell, m, n)$, the quantities $\delta \omega_{\ell m n} = \left( \omega'_{\ell m n} - \omega_{\ell m n} \right)/\omega_{\ell m n}$, and $\delta \tau_{\ell m n} = \left( \tau'_{\ell m n} - \tau_{\ell m n} \right)/\tau_{\ell m n} $ vanish for Kerr-like spacetimes having stationarity, axisymmetry, and circularity~\cite{Berti:2005ys, Krivan:1997hc, Dorband:2006gg, Giesler:2024hcr}. In other words, any deviation of $\{\delta \omega_{\ell m n},\, \delta \tau_{\ell m n}\}$ from zero will signify a violation of the Kerr symmetries. In particular, assuming the presence of stationarity and axisymmetry required to define a quasi-equilibrium state of the BH in the ringdown phase, we may refer the aforesaid test as a ``test of circularity'' and continue to do so for the rest of the paper\footnote{It is worth mentioning that recent years have seen a surge of interest in non-circular spacetimes, see 
for example Ref.~\cite{Ghosh:2024arw} and references therein.}.

At this point, we must emphasize that the aforesaid method to probe non-circularity is essentially an \textit{effective bottom-up} prescription valid solely when the non-GR deviations are tiny and so are $\{\delta \omega_{\ell m n},\, \delta \tau_{\ell m n}\}$. In general, loss of circularity may cause more severe modifications and even a complete breakdown of the ringdown model in Eq.~\eqref{ringdown}. In such cases, one is left with no other options than to perform a BH ringdown analysis from scratch and deduce the right model. Fortunately, the brilliant observational success of Kerr paradigm lends a strong support for the effectiveness of our prescription, which we shall discuss in details in the next section. 

In this spirit, we perform a thorough Bayesian analysis to probe possible non-circular deviations from the Kerr paradigm using GW ringdown observations. For this purpose, we consider a particularly loud GW event, namely \texttt{GW150914}, with high signal-to-noise ratio (SNR) in the ringdown phase \cite{LIGO1,LIGO2}. Our results clearly demonstrates that non-circular deviations are consistent with zero within experimental accuracy and future detectors with better ringdown SNRs and improved sensitivity will be able to constrain such deviations from Kerr symmetries more stringently. 

In summary, our analysis upholds the Kerr paradigm using an alternative yet novel scheme, namely the ``test of spacetime symmetries.'' Moreover, the above test of non-circularity has significant theoretical and observational implications. Circularity, which ensures the constancy of horizon angular velocity~\cite{Frolov:1998wf} and surface gravity in a theory-independent manner~\cite{Heusler}, makes our test a potential probe for horizon rigidity and the BH zeroth law as well. Interestingly, though the stationary and axisymmetric BHs beyond vacuum GR or in modified gravity may not be circular in general, it has been shown that effective gravity theories perturbatively connected to GR maintain circularity under specific conditions~\cite{Xie:2021bur}. However, relaxing these conditions can indeed produce non-circular solutions, as exemplified by the DHOST BHs \cite{Anson:2020trg, BenAchour:2020fgy, Anson:2021yli}. Thus, by constraining non-circular deviations from Kerr, our work also provides stringent limits on such possibilities.\\

\noindent{\bf{\em Models with non-circular deviations.}}
As a direct consequences of Kerr symmetries, the corresponding QNM equation gives rise to a pair of distinct modes for any fixed values of $\{\ell, m, n\}$~\cite{Berti:2005ys, Berti:2005ys, Krivan:1997hc, Dorband:2006gg, Giesler:2024hcr}. Let us now illustrate this for $\{2, \pm 2, 0\}$ modes, which are parameterized by $\{\omega_{2\pm20}^{(j)},\, \tau_{2\pm20}^{(j)}\}$. Here, the superscript $j = 1$ ($2$) represents the corresponding positive (negative) frequency mode. Though both the frequencies and damping times associated with $m=\pm 2$-modes are distinct for a fixed $j$, nevertheless there are relations such as: $\omega_{2-20}^{(j)}=-\omega_{220}^{(k)}$ and $\tau_{2-20}^{(j)}=\tau_{220}^{(k)}$ for $j \neq k$~\cite{Berti:2005ys, Krivan:1997hc, Dorband:2006gg, Giesler:2024hcr}. Therefore, it is sufficient to consider only the positive-frequency modes for both $\{2,\pm 2,0\}$ as independent parameters. However, in LVK observations of post-merger phases, the positive-frequency $\{2,-2,0\}$ mode is hardly excited for aligned-spin binaries~\cite{Berti:2005ys, Lim:2019xrb, Li:2021wgz}. As a result, the ringdown signal for the final Kerr BH is well described by,
\begin{equation} \label{Kerr}
    \begin{split}
        h_0(t) = &\, \mathcal{A}_{2 2 0}\, e^{\mathrm{i}\left(\omega_{2 2 0} t+\phi_{2 2 0}\right)}\, e^{-t / \tau_{2 2 0}}\, S_{2 2 0}(\theta)\, +\, \\
       &\mathcal{A}_{2 2 0}\, e^{-\mathrm{i}\left(\omega_{2 2 0} t+\phi_{2 2 0}\right)}\, e^{-t / \tau_{2 2 0}}\, S^*_{2 2 0}(\theta).
    \end{split}
\end{equation}
Here, we have fixed the azimuthal angle $\varphi = 0$, as it is degenerate with the modes’ initial phase $\phi$~\cite{Capano:2021etf}.
However, in the absence of circularity (deviation is assumed to be small), the above expression will be modified as
\begin{equation} \label{postKerr}
    \begin{split}
        h_1(t) = &\, \mathcal{A}_{2 2 0}\, e^{\mathrm{i}\left(\omega_{2 2 0} t+\phi_{2 2 0}\right)}\, e^{-t / \tau_{2 2 0}}\, S_{2 2 0}(\theta)\, +\, \\
       &\mathcal{A}_{2 2 0}\, e^{-\mathrm{i}\left(\omega'_{2 2 0} t+\phi_{2 2 0}\right)}\, e^{-t / \tau'_{2 2 0}}\, S^*_{2 2 0}(\theta),
    \end{split}
\end{equation}
where we consider $\omega'_{220}=\omega_{220}(1+\delta \omega_1)$ and $\tau'_{220}=\tau_{220}(1+\delta \tau_1)$. Here, the free parameters $\delta \omega_1$ and $\delta \tau_1$ signify possible departures from circularity in the $220$-mode. Our goal is to constraint these parameters from GW observations and check which model between $[{\cal H}_0:$ Kerr model given by Eq.~\eqref{Kerr}$]$ and $[{\cal H}_1:$ circularity-violation model given by Eq.~\eqref{postKerr}$]$ is preferred by the observed ringdown data. 

Due to the debated detection of overtones, we mostly concentrate on the fundamental mode and its possible non-circular departures. However, the accuracy of this model (${\cal H}_1$) in describing the ringdown GW signal forces us to consider a limited part of the observed signal restricted away from the merger (so that all other overtones get sufficiently damped). This reduces the overall post-merger SNR. On the other hand, including more overtones introduces new independent parameters, creating significant degeneracies among them and affecting the faithfulness of the model. Hence, as a balance between these two opposite cases, we consider another model ${\cal H}_2$ described by,
\begin{equation} \label{postKerr2}
    \begin{split}
        h_2(t) = &\, h_1(t) + \mathcal{A}_{2 2 1}\, e^{\mathrm{i}\left(\omega_{2 2 1} t+\phi_{2 2 1}\right)}\, e^{-t / \tau_{2 2 1}}\, S_{2 2 1}(\theta)\, +\, \\
       &\mathcal{A}_{2 2 1}\, e^{-\mathrm{i}\left(\omega'_{2 2 1} t+\phi_{2 2 1}\right)}\, e^{-t / \tau'_{2 2 1}}\, S^*_{2 2 1}(\theta)\, .
    \end{split}
\end{equation}
Now, we have additional parameters $\{\delta \omega_2, \delta \tau_2\}$ besides $\{\delta \omega_1, \delta \tau_1\}$, introduced by $\omega'_{221}=\omega_{221}(1+\delta \omega_2)$ and $\tau'_{221}=\tau_{221}(1+\delta \tau_2)$. However, the inclusion of the $221$-mode will help us better represent the prompt ringdown signal with more accuracy. Note that, in constructing both of the above models, we have assumed the perturbation has the equatorial reflection symmetry like the Kerr case~\cite{Blanchet:2013haa, Capano:2021etf}. It fixes the amplitudes and phases of various modes unambiguously. This assumption is not in contradiction with our non-circular construction, since the breaking of circularity does not necessarily break the equatorial $Z_2$ symmetry of the perturbations. However, for the most general scenario, departures from Kerr symmetry could induce a breaking of equatorial $Z_2$ symmetry as well.\footnote{We thank Vitor Cardoso for pointing this out to us.} We have conducted an exhaustive study of this scenario in Appendix B and have not seen any significant difference from our present results.

A few important comments are in order. One must appreciate the difference between our models and that for the no-hair test~\cite{Isi:2019aib}. The post-Kerr paradigm can be approached in two distinct ways. The first involves introducing a violation of the no-hair theorem by adding new hairs to the Kerr BH, while preserving stationarity, axisymmetry, and circularity. This is the method typically employed in tests of the no-hair theorem~\cite{Isi:2019aib}. In our model, however, we focus on modifying Kerr symmetries that induce a relative shift among the twin modes and identifying observational signatures of such modifications. Hence, we have set the additional hairy parameters, which fail to produce such relative shift and are tightly constrained already in Ref.~\cite{Isi:2019aib}, to zero. Consequently, in our framework, vanishing deviation parameters serve as a necessary and sufficient condition for the preservation of Kerr isometries, independent of the specifics of the new gravitational physics. Thus, our approach provides a novel, \textit{theory-independent} new test of the Kerr paradigm.\\

\begin{figure*}[!tb]
\centering
\includegraphics[width=0.49\textwidth]{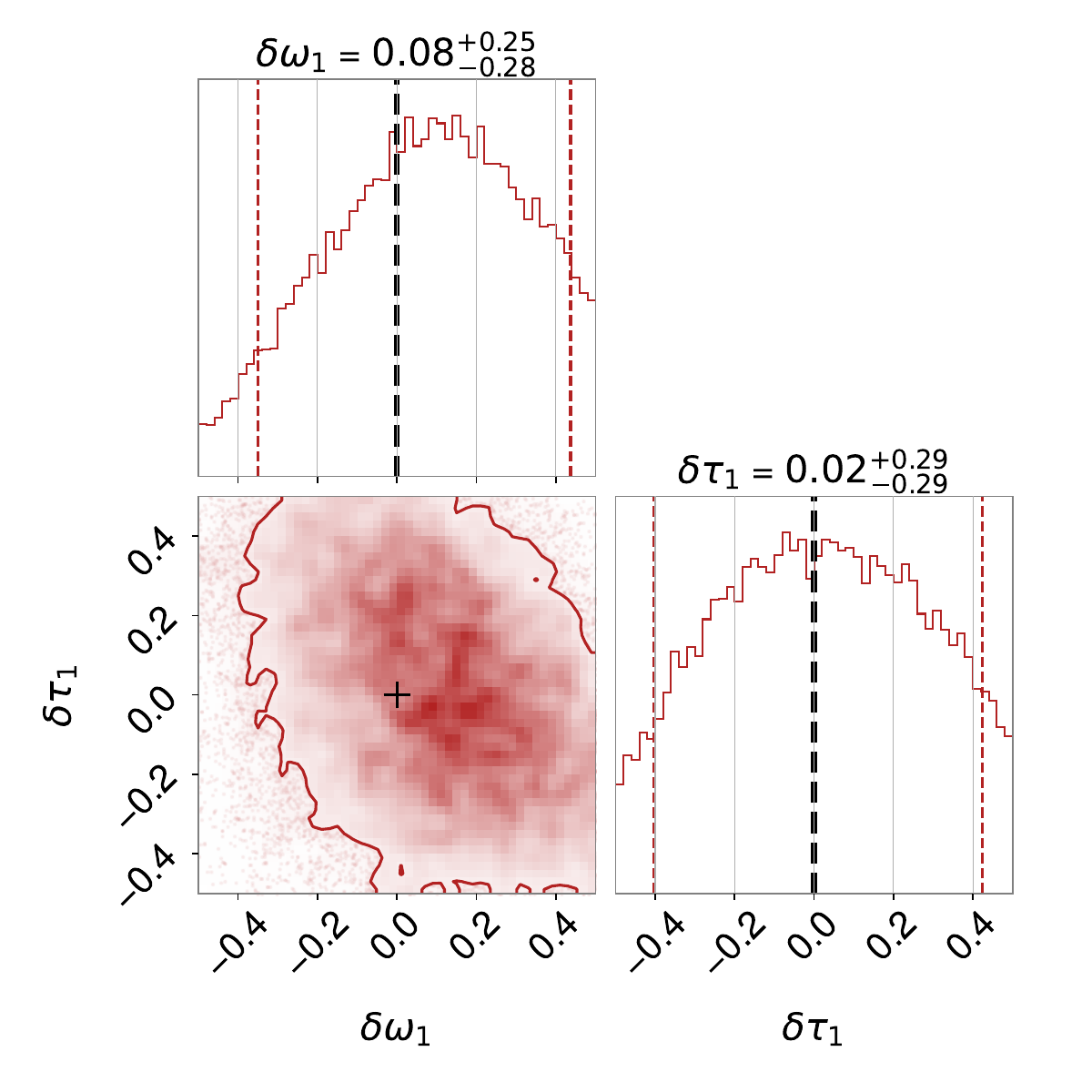}
\includegraphics[width=0.49\textwidth]{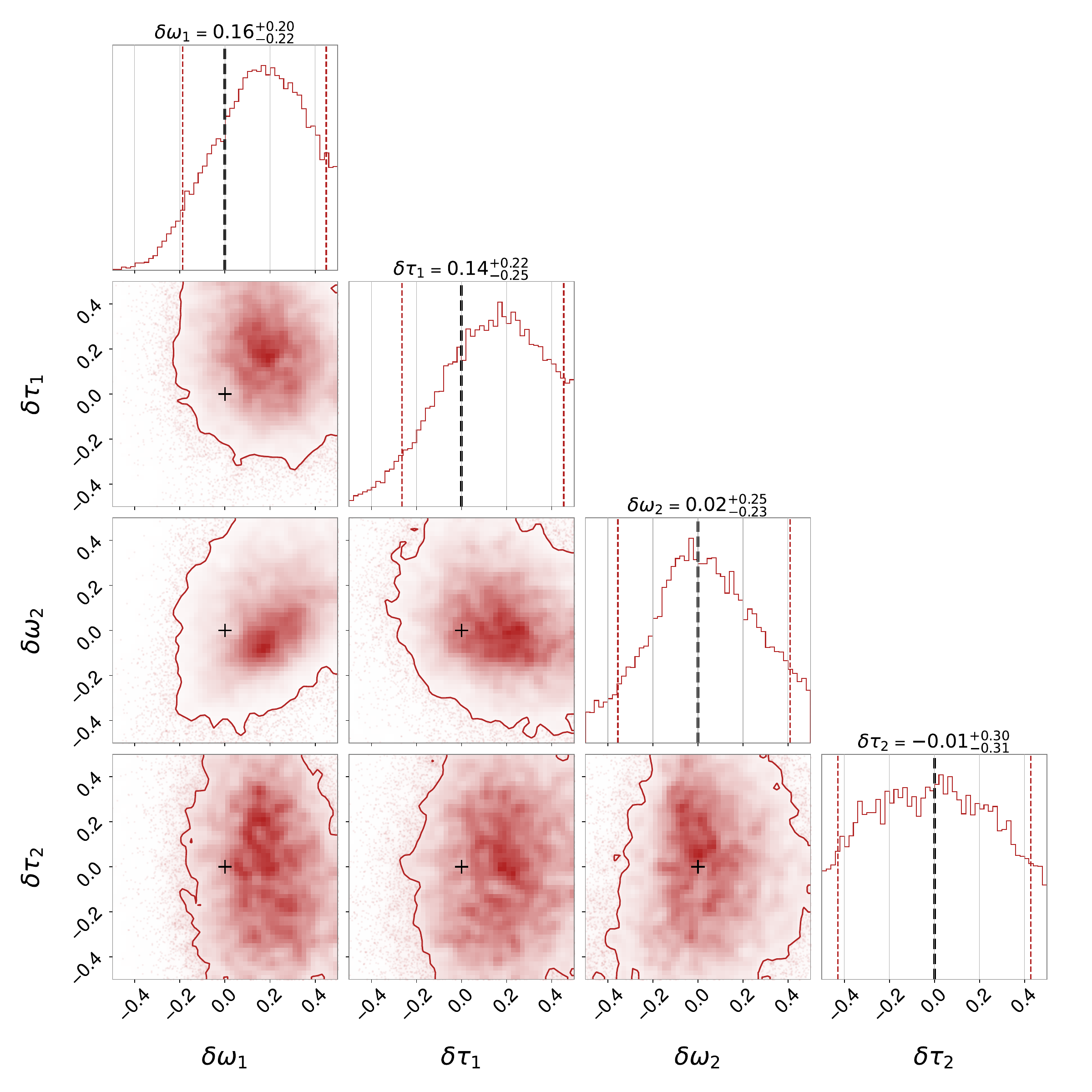}

\caption{Marginalized posterior distribution of the non-circularity parameters obtained from \texttt{GW150914} ringdown data corresponding to the model ${\cal H}_1$ (left panel) and ${\cal H}_2$ (right panel). The contours indicate the $90\%$ credible regions, while the black lines and plus signs mark the null hypothesis.
}
\label{GW150914_10mf_N0_plot}
\end{figure*}
\noindent{\bf{\em Parameter Estimation with \texttt{GW150914}.}} We consider the models presented in Eq.~\eqref{postKerr} and Eq.~\eqref{postKerr2} to conduct a comparative Bayesian analysis with the Kerr-model given by Eq.~\eqref{Kerr} using the ringdown data for the event \texttt{GW150914}. The analysis duration is set to 0.2 seconds from the ringdown start time ($t_0$), which for model ${\cal H}_1$ is taken to be $10\, M_{f}$ after the peak time ($1126259462.423$ GPS for the LIGO Hanford detector~\cite{LIGOScientific:2016lio}). In contrast, for the model that includes an overtone (${\cal H}_2$), the ringdown start time is set at the merger itself. This adjustment accounts for the contribution from the overtone which is expected to be dominant immediately after the merger. We also time shift the LIGO Livingston data to account for the arrival-time delay and align the signals from the two detectors. For computational simplicity, we work with fixed sky locations $(\alpha, \delta) = (1.95, -1.27)$. We apply uniform prior on all search parameters: $M_f \in [20, 120] M_\odot$, $\chi_f \in [0.0, 0.99]$, $A_{22n} \in [0.0, 2.5\times 10^{-19}]$, $\phi_{22n} \in [0.0, 2\pi]$, $(\iota, \psi) \in [0, \pi]$, and $(\delta \omega_{n}, \delta \tau_{n}) \in [-0.5, 0.5]$. With the likelihood and prior functions we perform parameter estimation with the \texttt{DynamicNestedSampler} from \texttt{dynesty}~\cite{Speagle_2020, dynesty}, within the time-domain Bayesian analysis framework developed in Ref. \cite{Mishra:2023kng}. The sampling was performed with the ``unif'' method, using $3000$ live points while the remaining sampler parameters are kept at their default settings. For more details, see Appendix A.

In Fig.~[\ref{GW150914_10mf_N0_plot}] we present the corner plots, illustrating  the marginalized posterior distribution of the deviation parameters for models ${\cal H}_1$ (left) and ${\cal H}_2$ (right). For the full corner plots, readers are referred to Fig.~[\ref{GW150914_10mf_N0_plot_full}] and Fig.~[\ref{GW150914_overtone}] in Appendix B. Notably, the final mass and final spins show degeneracy (degree of degeneracy depend on the model under consideration) with the fractional deviations in frequency and decay time. This degeneracy springs from the nature of the ringdown analysis itself. For ${\cal H}_1$, we extract two observable data points, namely the frequency and decay time, while attempting to estimate four unknowns: final mass, final spin and two deviation parameters. The overtone model ${\cal H}_2$ further introduces additional parameters: an amplitude, phase, and two fractional deviation parameters for the overtone frequency and decay time. The null hypothesis $(\delta \omega_{n} = \delta \tau_{n}=0)$ lies well within the $90\%$ credible region for both the models, suggesting no significant deviation from circularity. 
This statement is further reinforced by our computations of the Bayes factor by the Savage-Dickey density ratio~\cite{Dickey1971}:
\begin{equation}
    \label{BayesFactor}
    {\mathcal{B}}_{{\cal H}_a}^{{\cal H}_0} = \frac{p(\vec d \mid \mathcal{H}_0)}{p(\vec d \mid \mathcal{H}_a)} = \frac{p(\{\delta \omega_n, \delta \tau_n\}=0 \mid \vec d, \mathcal{H}_a)}{p(\{\delta \omega_n, \delta \tau_n\} \mid  \mathcal{H}_a)}\, , \nonumber
\end{equation}
where $\mathcal{H}_0$ ($\mathcal{H}_a$) is the null (alternative) hypothesis. For our non-circular models, Bayes factors are $\mathcal{B}^{\rm Kerr}_{{\cal H}_1} = 2.1$  and  $\mathcal{B}^{\rm Kerr}_{{\cal H}_2} = 2.4$, respectively. As per Jeffrey's interpretation~\cite{Jeffreys:1939xee}, these evidences are ``worth mentioning'' in favor of the circular Kerr paradigm.\\

\begin{figure}[!tb]
\centering
\includegraphics[width=\columnwidth]{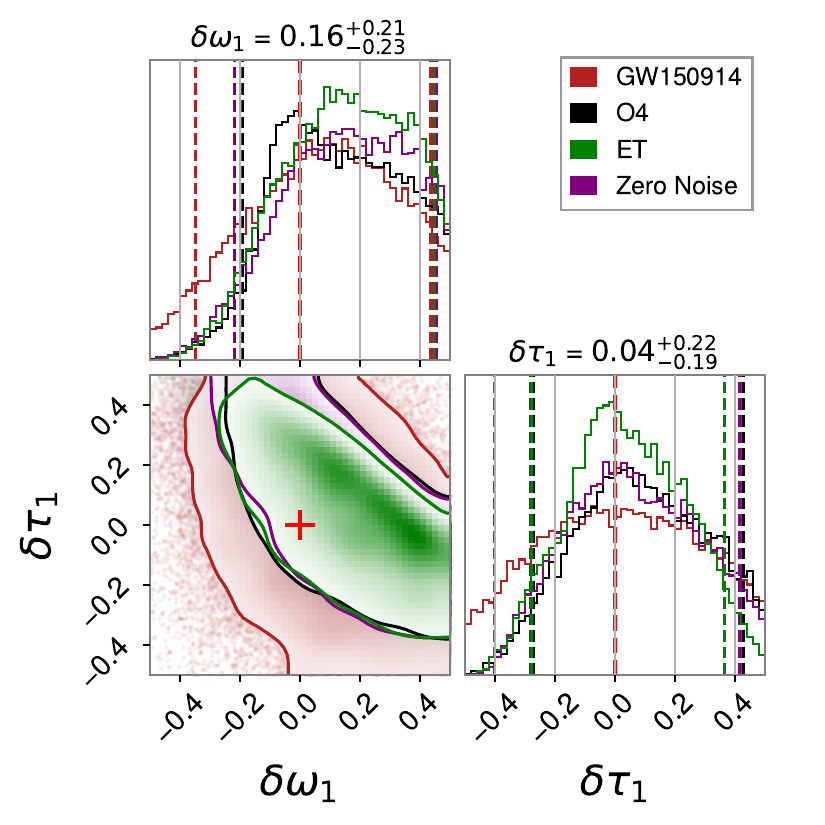}
\caption{Marginalized posterior distribution for the non-circularity parameters of a \texttt{GW150914}-like ringdown signal, detected with ET (green) and O4 (black) sensitivity, is shown. For comparison, we also include the results from the analysis of \texttt{GW150914} and a zero-noise injection. The contours indicate the $90\%$ credible regions, while the dashed red lines and plus sign mark the null hypothesis.}
\label{future_obs}
\end{figure}

\noindent{\bf{\em Future Observations.}} Future generation detectors such as the Einstein Telescope (ET)~\cite{einstein_telescope} and the Laser Interferometer Space Antenna (LISA)~\cite{lisa} will have much better sensitivity and improved SNR compared to the current LVK detectors. Such improvements in SNR are particularly crucial for ringdown analysis and leads to a reduction in the overall posterior volume. However, even with higher SNR, it may not be sufficient to break the degeneracy among various aforementioned parameters. In this section, we simulate a \texttt{GW150914}-like ringdown injection using ${\cal H}_1$ template with the current noise sensitivity of O4 and improved sensitivity of future detectors like ET (ET-D sensitivity curve~\cite{et_d_sensitivity}). In addition, we also perform a zero-noise injection analysis (with O4 sensitivity curve) for comparison. The remnant parameters used for the injection are as follows: $M_f = 70\, M_\odot, \chi_f = 0.67, A_{220} = 9.45 \times 10^{-21}, \phi_{220} = 3.29, \iota = \pi, \alpha = 1.95, \delta = -1.27, \psi = 0.82, \delta \omega_{1} =  \delta \tau_{1} = 0.0$. This yields a SNR of  $\sim 34$ for O4 and $156$ for ET. In Fig. [\ref{future_obs}] we illustrate the recovery of the deviation parameters using the above injection setup. Additionally, Fig.~[\ref{future}] in Appendix B shows the full posterior distributions for all the search parameters, providing a comprehensive overview of the injection and recovery across different sensitivity. \\

\

\noindent{\bf{\em Discussion and Conclusions.}}
The uniqueness theorems of GR dictates the Kerr spacetime as the unique stationary and asymptotically flat BH solution to Einstein field equations in vacuum. These theorems are also tied to rigidity theorems, underscoring the remarkable symmetries of Kerr BHs. Recently, these profound results have been extended to higher-curvature gravity theories, provided the solutions are perturbatively connected to those in GR (for both vacuum and non-vacuum cases)~\cite{Xie:2021bur}. Consequently, any deviation from the Kerr symmetries in a BH spacetime would represent a radical departure from the metric paradigm of gravity. For instance, such departures could emerge from non-minimal couplings or suitable non-metric gravitational degrees of freedom, as seen in DHOST theories~\cite{Anson:2020trg, BenAchour:2020fgy, Anson:2021yli}. Alternatively, such breakdown of Kerr symmetries can also occur if the compact object is not a BH but rather a horizonless mimicker. 

Given the significance of such possibilities, it is crucial to pursue observational searches for deviations from Kerr symmetries and to impose constraints on these departures. GW observations offer a powerful avenue for conducting these tests. However, we still lack a complete understanding of various GW signatures for an exact post-Kerr BH solution lacking Kerr-like isometries. Therefore, a phenomenological approach becomes indispensable, allowing us to model the key features that could emerge from violations of Kerr symmetries.

Motivated by the connection between symmetry and degeneracy, we adopt well-motivated waveform models for the post-merger ringdown signal and conduct a detailed Bayesian estimation of the deviation parameters. Additionally, we examine the impact of incorporating higher overtones into our analysis and explore the potential for constraining these parameters using future detectors. In all cases, we observe that the current GW observations are entirely consistent with Kerr symmetries within $90\% $ credibility. An extension of our work could involve studying the ringdown signature of a specific gravity theory (like DHOST) with a rotating BH solution breaking some Kerr symmetries, and examining how our constraints lead to bounds on its parameters.

In conclusion, the Kerr paradigm remains central to our understanding of compact objects in GR, with its symmetries originating from uniqueness and rigidity theorems. Our phenomenological approach using GW observations suggests that current observations are consistent with such Kerr symmetries, and future tests could further refine these constraints, potentially identifying new boundaries for theories beyond GR.\\

\noindent{\bf{\em Acknowledgments.}}  We extend our gratitude to the members of the ``Testing General Relativity'' group within the LIGO-Virgo-KAGRA collaboration, as well as the Astrophysical Relativity group at ICTS, for their valuable insights and engaging discussions. We sincerely thank Vitor Cardoso, Gregorio Carullo, Enrico Barausse, and N. V. Krishnendu for reviewing the draft and providing valuable feedback. The computations were conducted using the Sonic HPC cluster at ICTS-TIFR, utilizing publicly available software packages such as \texttt{dynesty}~\cite{Speagle_2020, dynesty}, \texttt{pyRing}~\cite{pyring}, \texttt{lalsuite}~\cite{lalsuite}, \texttt{matplotlib}~\cite{matplotlib}, \texttt{numpy}~\cite{numpy}, \texttt{scipy}~\cite{scipy}, and \texttt{corner}~\cite{corner}. This material is based upon work supported by NSF's LIGO Laboratory which is a major facility fully funded by the National Science Foundation. The research of SS is supported by the Department of Science and Technology, Government of India, under the ANRF CRG Grant (No. CRG/2023/000934). 
\onecolumngrid
\appendix
\section{Appendix A. Analysis Details} \label{analysis}
Conventional LIGO data analysis is typically performed in the frequency domain using full inspiral-merger-ringdown (IMR) waveform models. However, when analyzing only a specific segment of the signal, standard frequency domain methods may not be applicable due to spectral leakage caused by abrupt data cuts. For short-duration signals like the ringdown, this can significantly distort the frequency domain representation. To avoid this, we adopt a time-domain approach which eliminates the need for a Fourier transform to the frequency domain but introduces a non-diagonal likelihood defined as~\cite{Isi:2021iql}, 

\begin{equation}\label{TD_likelihood}
\ln\, \mathcal{L}(d|\theta) = -\frac{1}{2}\sum_{i,j = 0}^{N-1}\left[d_i - h_i(\theta)\right] C_{i j}^{-1} \left[d_j - h_j(\theta)\right]~. \nonumber
\end{equation}

Here, N represents the number of data points while $\theta$ encompasses all the intrinsic and extrinsic parameters of the system. For a network of detectors, the joint likelihood is obtained by multiplying the individual likelihoods from each detector. The ringdown data is expressed as $d(t) = h(t) + n(t)$, where $h(t)$ represents the ringdown waveform as observed by each individual detectors,

\begin{equation}\label{response}
    h(t) = h_{+}(t)\, F_{+}(\alpha, \delta, \psi) + h_{\times}(t)\, F_{\times}(\alpha, \delta, \psi)\, , \nonumber
\end{equation}
where $F_{+}$ and $F_{\times}$ are the detector response function (computed using the \texttt{lalsuite} library~\cite{lalsuite}) which depends on the right ascension ($\alpha$), declination($\delta$) and the polarization angle ($\psi$).  Moreover, $n(t)$ denotes the noise component which is well-approximated by a wide-sense stationary Gaussian process, hence fully characterized by a symmetric Toeplitz covariance matrix:
\begin{equation}
C_{ij} = \rho(|i-j|) , \quad\quad  \rho(k) \propto \sum_i n_{i}\, n_{i + k}~, \nonumber
\end{equation}
where $\rho(k)$ is the auto-correlation function (ACF). In the data preprocessing stage, we band-pass $4096$ seconds of data, sampled at $4096$ Hz, around the merger time using a fourth-order Butterworth filter in the range $[20, 2047]$ Hz. The ACF is computed using the FFT method implemented in \texttt{pyRing}~\cite{pyring}. In terms of the above likelihood function and prior $\pi(\theta)$, the posterior distribution can be expressed as
\begin{equation}
p(\theta|d)=\frac{\mathcal{L}(d|\theta)\;\pi(\theta)}{\int d\theta \mathcal{L}(d|\theta)\,\pi(\theta)}\, . \nonumber
\end{equation}

\section{Appendix B. Full Corner Plots} \label{figures}
\begin{figure*}
\centering
\includegraphics[width=0.49\textwidth]{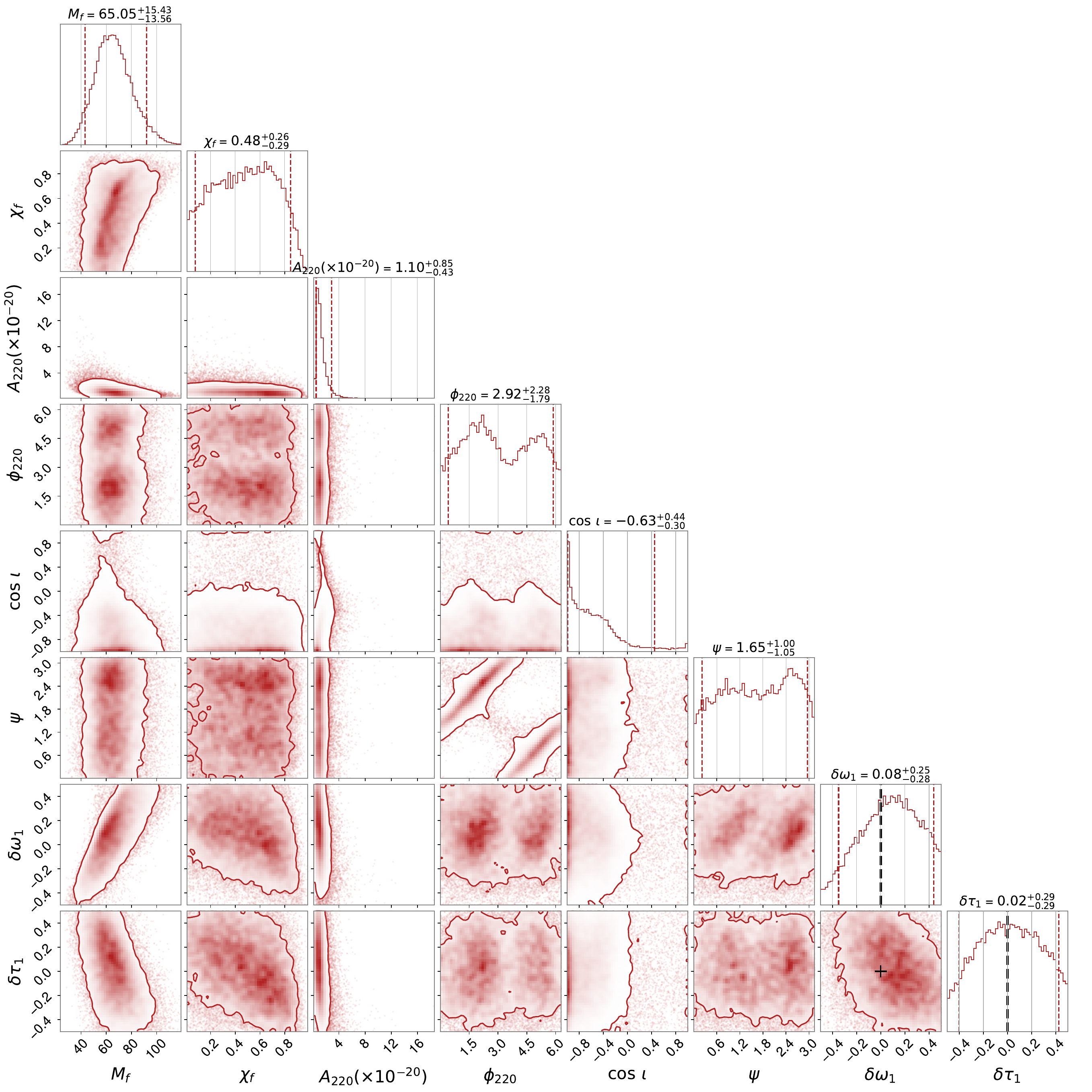}
\includegraphics[width=0.49\textwidth]{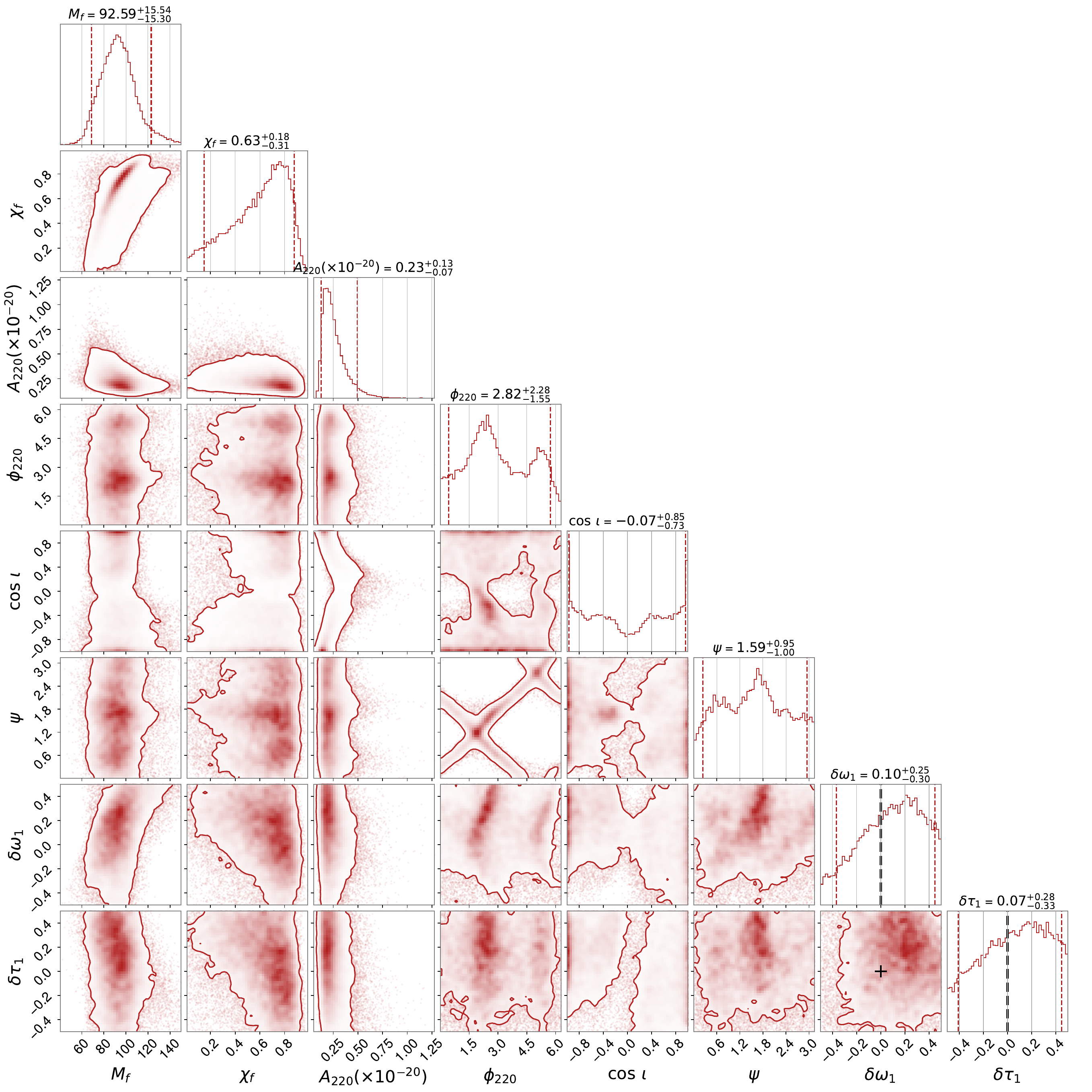}

\caption{In this figure, we present the full corner plot for all the search parameters corresponding to model ${\cal H}_1$ for the events \texttt{GW150914} (left panel) and \texttt{GW190521\_074359} (right panel). The contours indicate the $90\%$ credible regions, while the black lines and plus signs mark the null hypothesis.}
\label{GW150914_10mf_N0_plot_full}
\end{figure*}

Fig.~[\ref{GW150914_10mf_N0_plot_full}] illustrates the posterior  distribution involving all the search parameters for the model ${\cal H}_1$ corresponding to the events \texttt{GW150914} and \texttt{GW190521\_074359}~\cite{LIGOScientific:2020ibl}. For the event \texttt{GW190521\_074359}, the ringdown start time is assumed to be $5\, M_{f}$ after the merger to ensure sufficient SNR in the analysis segment. The prior on the final mass is taken to be $M_f \in [20, 150] , M_\odot$, while the sky location parameters are fixed at their maximum-likelihood values obtained from a full IMR analysis. 

\begin{figure*}[!tb]
\centering
\includegraphics[width=0.99\textwidth]{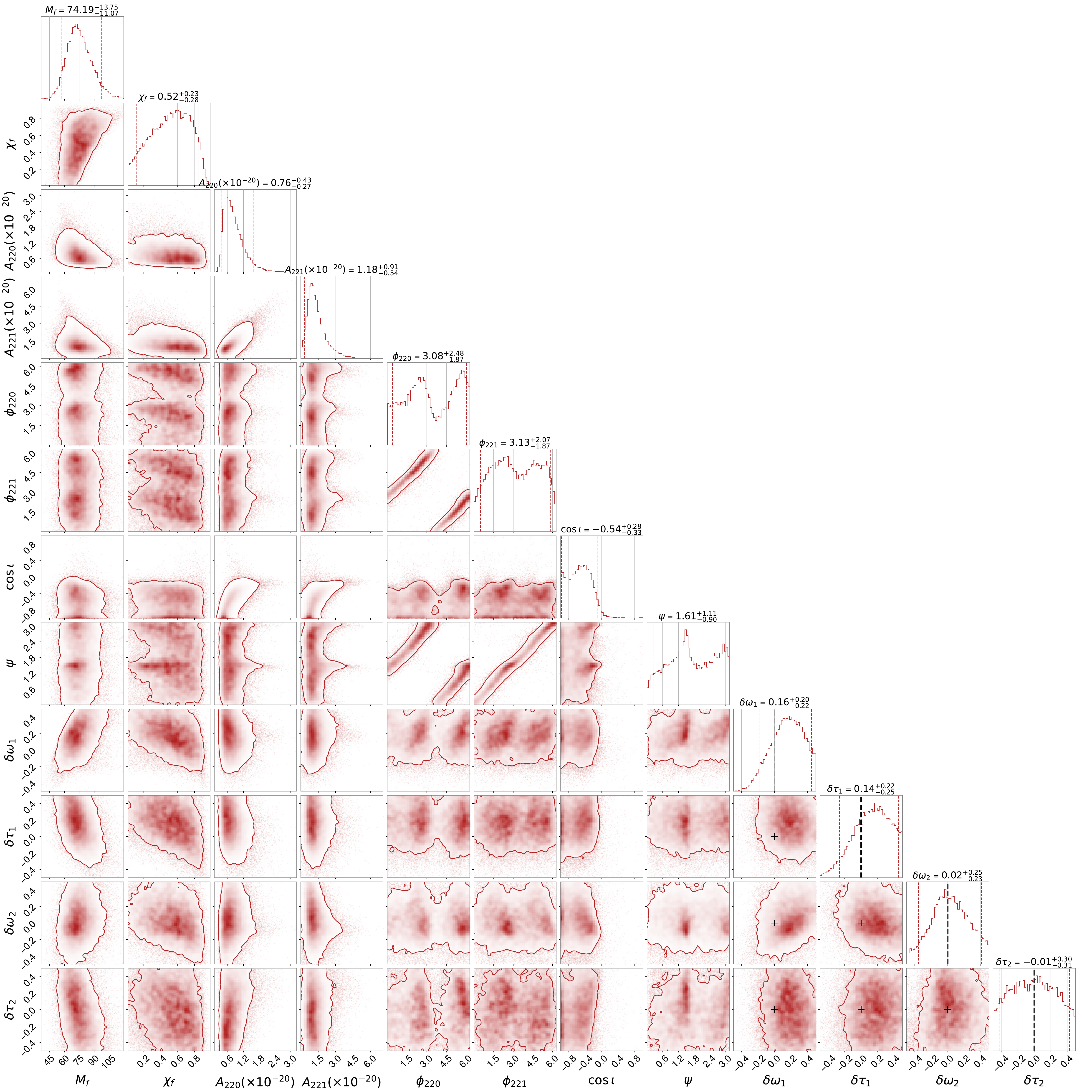}

\caption{The posterior distribution for all search parameters derived from the \texttt{GW150914} ringdown data, using the overtone model ${\cal H}_2$, highlighting the correlations between different parameters. The contours represent the $90\%$ credible regions, with the black lines and plus signs indicating the null hypothesis.
}
\label{GW150914_overtone}
\end{figure*}

Likewise, Fig.~[\ref{GW150914_overtone}] displays the posterior distributions for different search parameters corresponding to the overtone model ${\cal H}_2$ for the event \texttt{GW150914}. The circularity deviation parameter in frequency ($\delta \omega_n$) exhibits a positive correlation with the remnant mass, while the deviation in damping time ($\delta \tau_n$), shows a negative correlation. However both ($\delta \omega_n$, $\delta \tau_n$) are negatively correlated with the remnant spin, leading to an effectively lower inferred measurement of the spin. Furthermore, we studied the measurement prospects of these additional parameters in reference to current and future generation detectors. 

\begin{figure*}[!tb]
\centering
\includegraphics[width=0.8\textwidth]{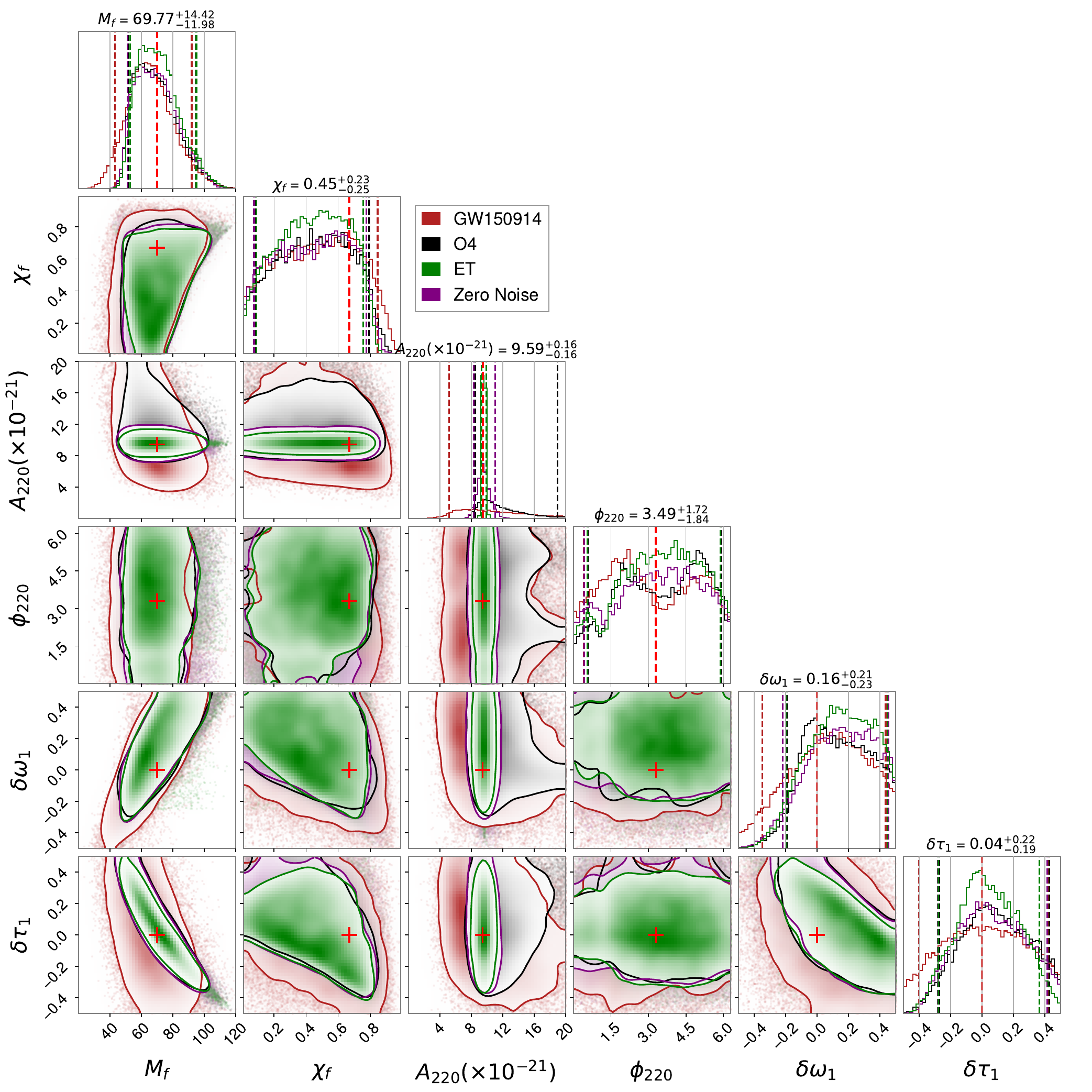}

\caption{Future Observations: In this figure, we show the full corner plot for all the search parameters corresponding to model ${\cal H}_1$ for a \texttt{GW150914}-like ringdown injection, using the O4 and ET noise sensitivity curves. For comparison, we also include the results from the analysis of \texttt{GW150914} and a zero-noise injection. The vertical lines corresponds to the respective $90\%$ credible regions, while the plus signs marks the injected values.}
\label{future}
\end{figure*}

In Fig.~[\ref{future}], we demonstrate this in the context of O4 and Einstein Telescope detector sensitivity. Improved noise sensitivity leads to better measurements of frequency and decay time, resulting in a reduced posterior volume. However, this does not necessarily eliminate degeneracies between the parameters.

\begin{figure*}[!tb]
\centering
\includegraphics[width=0.8\textwidth]{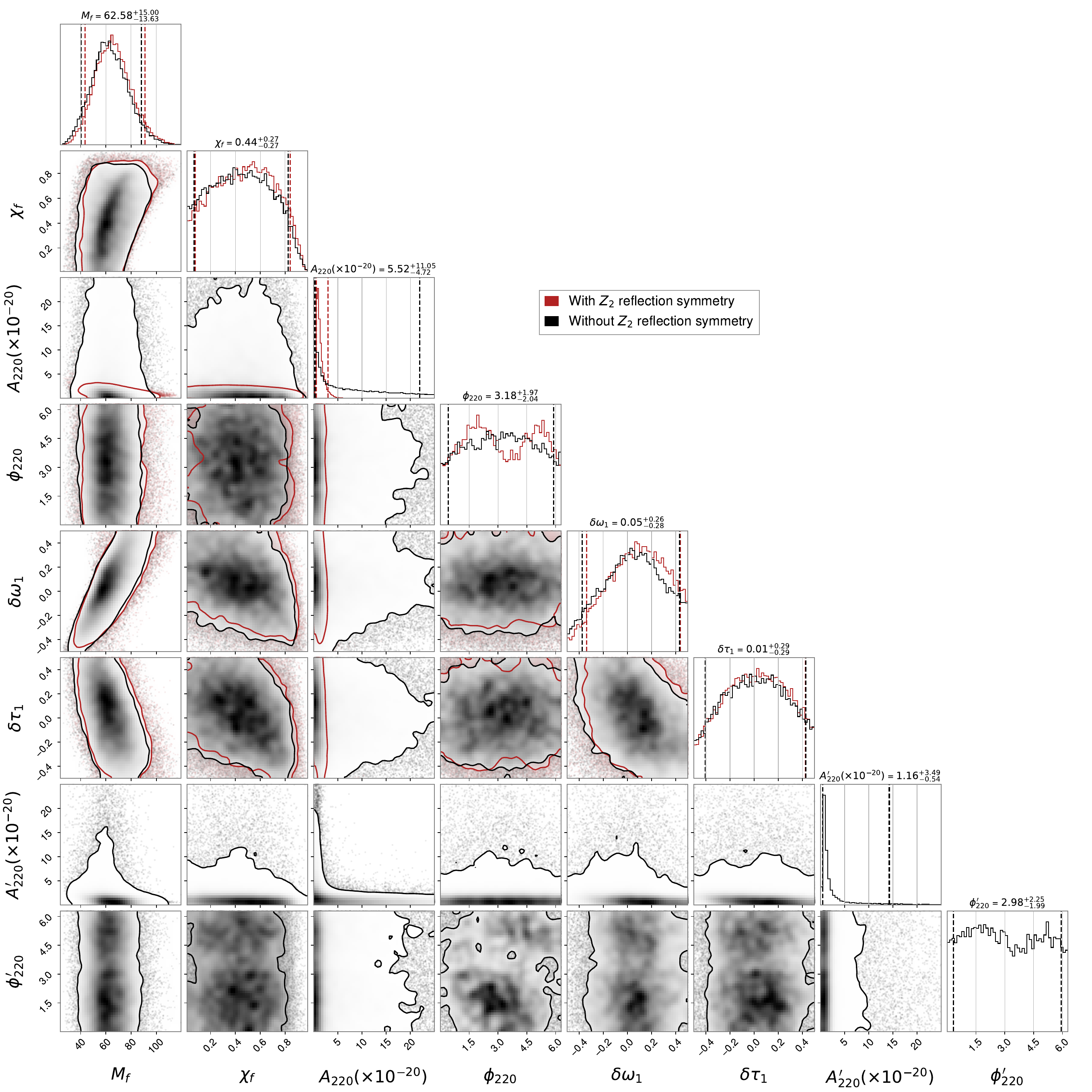}

\caption{This figure presents a comparative analysis of the posterior distributions for the search parameters associated with the non-circular ringdown model, both with and without equatorial reflection symmetry for \texttt{GW150914}.}
\label{eqt_vs_noneqt}
\end{figure*}

We have also studied the scenario where a breaking of Kerr symmetry might induce a departure from the equatorial $Z_2$ symmetry of the the perturbations. Although such a scenario is not necessary to happen, but nevertheless we should be open for its occurrence in the most general case. For this purpose, we performed an analysis incorporating different amplitudes and phases for the twin modes, alongside non-circularity parameters. A comparative posterior plot is shown in Fig.~[\ref{eqt_vs_noneqt}] for the following cases:
(i) Non-circular but equatorial $Z_2$-symmetric model given in Eq. (3) of the main text with parameters $\{\mathcal{A}_{220} = \mathcal{A}'_{220}, \phi_{220} = \phi'_{220}, \delta \omega_1, \delta \tau_1\}$ shown in red. 
(ii) New non-circular and no equatorial $Z_2$ symmetry model with parameters $\{\mathcal{A}_{220},  \mathcal{A}'_{220}, \phi_{220}, \phi'_{220}, \delta \omega_1, \delta \tau_1\}$ shown in black. Note that in the later case, the amplitudes and phases of the twin modes do not obey the relation $\mathcal{A}'_{220}\, e^{\mathrm{i} \phi'_{220}} = \mathcal{A}_{220}\, e^{-\mathrm{i} \phi_{220}}$ in the absence of the equatorial reflection symmetry. From the figure, it is evident that there is no significant difference in the posteriors of $\{\delta \omega_1, \delta \tau_1\}$ for the two aforementioned cases. The only notable change is an increase in the $90\%$ CI for $A_{220}$ compared to the equatorial $Z_2$-symmetric case. Moreover, to quantify the similarity between the posteriors, we calculated the Kullback–Leibler (KL) divergences. For $\delta \omega_1$, $D_{KL} = 0.01$ and for $\delta \tau_1$, $D_{KL} = 0.008$.

Moreover, a straightforward mismatch analysis with improved SNRs clearly demonstrates the distinguishability between the Kerr model in Eq.~\eqref{Kerr} and the non-circular model in Eq.~\eqref{postKerr} even with small non-circular deviations. For example, a $\texttt{GW150914}$-like signal with deviations $(\delta \omega_1, \delta \tau_1) = (0.04, 0.04)$ results in a mismatch exceeding $6\%$ at ET sensitivity (similar for CE). However, we note that the actual recovery of these deviation parameters might pose a few challenges due to their degeneracies with other intrinsic parameters (see the Appendix B). But this feature is not unexpected in such post-Kerr tests. We target to perform a thorough analysis of such features in a future work.


\twocolumngrid

\end{document}